\documentclass[aps, prd, superscriptaddress, twocolumn, nofootinbib]{revtex4-1}

\usepackage[colorlinks, linkcolor=red, anchorcolor=green, citecolor=blue]{hyperref}
\usepackage{amsmath}
\usepackage{amssymb}
\usepackage{graphicx}
\usepackage{float}
\usepackage{color}
\usepackage{braket}
\usepackage{orcidlink}
\usepackage{comment}
\graphicspath{{figures}}
\allowdisplaybreaks[4]

\newcommand{\bp}{\boldsymbol{p}}
\newcommand{\bk}{\boldsymbol{k}}

\begin{document}

\title{Probing Pair Correlations in QCD Matter with Photon Spectra}

\author{Xingjian Lu}
\email[]{lus21@mails.tsinghua.edu.cn}
\affiliation{Department of Physics, Tsinghua University, Beijing 100084, China}

\author{Shuzhe Shi}
\email[]{shuzhe-shi@tsinghua.edu.cn}
\affiliation{Department of Physics, Tsinghua University, Beijing 100084, China}

\begin{abstract}
Correlations in the phase-space distribution of partons play an important role in the initial stage of relativistic heavy-ion collisions, where the matter is dense and far from equilibrium. Photons produced in the hot medium, which predominantly originate from two-parton initial states, are sensitive to two-particle correlations in the phase-space distribution. In this work, we study how pair correlations in non-equilibrium QCD matter affect in-medium photon production. We decompose the two-particle distribution as $\mathcal F_{ab}=f_a f_b+g_{ab}$, where $g_{ab}$ is the pair correlation. Focusing on the $2\to2$ quark--antiquark annihilation and Compton channels, we compute the leading-logarithmic photon spectrum by expanding the single-particle distribution and pair correlation in a spectral basis, thereby accommodating a broad class of two-particle distributions. For a rotationally invariant medium, we find that relative-angle modes of the pair correlation generate sign-changing modifications to the photon spectrum, with magnitudes that can be comparable to the factorized contribution. Thus, photon spectra, although single-particle observables, can measure the momentum correlations of the emitting medium and therefore probe the early-time hydrodynamization.
\end{abstract} 

\maketitle
\section{Introduction}
The study of electromagnetic radiation from strongly interacting matter has long provided an important window into the properties of hot and dense quantum chromodynamics (QCD) matter~\cite{Feinberg:1976ua, Shuryak:1978ij, Kajantie:1981wg, McLerran:1984ay, Kajantie:1986dh, Tserruya:2009zt}. See~\cite{Gale:2025ome, Jackson:2025mtk, Du:2025ith} for recent reviews and~\cite{Wu:2024vyc,Gao:2026vxs} for related recent studies. Relativistic heavy-ion collisions create a transient medium in which quarks and gluons are expected to be liberated from hadrons and to form a deconfined state of matter. Since the lifetime and spatial extent of this medium are limited, its properties must be inferred from particles emitted during and after its evolution. Among the available observables, photons and dileptons are particularly valuable because they are not subject to strong final-state interactions~\cite{PHENIX:2008uif, PHENIX:2011oxq, Liu:2012ax, Rapp:2014hha, ALICE:2018dti, Dasgupta:2020orj, Sun:2023rhh}. Once produced, they escape the medium almost undisturbed and thus retain information about the local environment in which they were emitted.

Photons measured in relativistic heavy-ion collisions receive contributions from several sources. The inclusive photon yield contains a large background from hadron decays~\cite{ALICE:2015xmh,Hui:2000yc}, dominated by the electromagnetic decays of neutral mesons, in particular
$
\pi^0 \to \gamma+\gamma,
\eta \to \gamma+\gamma .
$
After subtracting this decay component, the remaining photons are commonly referred to as direct photons,
$
  N^\gamma_{\rm direct}
  =
  N^\gamma_{\rm inclusive}
  -
  N^\gamma_{\rm decay}.
$
Direct photons, however, do not originate from a single production mechanism~\cite{Stankus:2005eq, Liu:2007tw, Vitev:2008vk, David:2019wpt}. They include prompt photons produced in initial hard partonic scatterings~\cite{Owens:1986mp}, photons emitted during the pre-equilibrium stage~\cite{Churchill:2020uvk, Garcia-Montero:2023lrd, Garcia-Montero:2024lbl}, thermal photons radiated from the near-equilibrium quark-gluon plasma (QGP)~\cite{Paquet:2015lta}, photons radiated from the hadronic gas~\cite{Turbide:2003si}, and possible photons associated with jet-medium interaction~\cite{Fries:2002kt, Turbide:2005fk, Qin:2009bk}. Among these contributions, photons emitted from the medium are of particular interest because they are sensitive to the microscopic details of the matter created in the collision~\cite{Paquet:2015lta}. In this work, we use the term \textit{in-medium photons} to refer to photons produced by microscopic interactions involving constituents of the QCD medium. In-medium photons provide a way to probe microscopic medium properties, such as non-equilibrium single-particle distributions and pair correlations, through their production rates. We therefore focus on in-medium photons rather than on the full experimentally measured direct-photon yield.

The in-medium photon yield is obtained by integrating the local photon emission rate over the space-time evolution of the medium,
\begin{equation}
  E\frac{dN^\gamma}{d^3p}
  =
  \int d^4x\,
  E^\ast \frac{dR^\gamma}{d^3p^\ast}.
\end{equation}
Here $E^\ast dR^\gamma/d^3p^\ast$ denotes the photon emission rate in the local rest frame of the medium. The space-time integral evaluates the local photon production rate throughout the evolving medium, using macroscopic fluid information such as the local temperature and flow velocity~\cite{Liu:2009kta,Dion:2011pp,Shen:2013vja,Gale:2014dfa,vanHees:2014ida,Jia:2022awu}. In the present study, however, we do not focus on the photon yield observed in the laboratory frame. Instead, we study the local in-medium photon production rate. This allows us to isolate how the microscopic properties of the emitting medium affect photon production before uncertainties from the global space-time evolution are introduced.

In-medium photons are produced from microscopic mechanisms including quark--antiquark annihilation, $q+\bar q \to g+\gamma$, and Compton scattering, $q(\bar{q})+g \to q(\bar{q})+\gamma$~\cite{Kapusta:1991qp, Baier:1991em}, as well as collinearly enhanced bremsstrahlung and inelastic pair annihilation --- in 2001, Arnold, Moore, and Yaffe showed that the latter contribute at the same order in the coupling~\cite{Arnold:2001ms, Arnold:2001ba}. Moreover, multiple scatterings in the medium modify these collinear processes through the Landau--Pomeranchuk--Migdal (LPM) effect~\cite{Landau:1953um, Landau:1953gr, Migdal:1956tc, Aurenche:2000gf, Arnold:2001ms}. Consequently, a complete leading-order calculation of the photon production rate requires the resummation of these LPM-suppressed collinear emissions, as established in the Arnold--Moore--Yaffe calculation~\cite{Arnold:2001ba, Arnold:2001ms, Arnold:2002ja}. In addition to the standard thermal rates, photon production has been studied in several extensions that emphasize different microscopic sources of electromagnetic radiation, including multiple rescatterings of an energetic quark jet~\cite{Zhang:2010hiv}, magnetic-field-induced emission~\cite{Sun:2023pil}, and production in a medium with a nontrivial Polyakov loop~\cite{Hidaka:2015ima}. The relation between the spectral shape of produced probes and the underlying scattering kernel has also been systematically analyzed~\cite{Lu:2026gcd}.

\vspace{3mm}

Despite extensive studies of photon production, the role of pair correlations in the medium remains less understood. Standard thermal rates usually use factorized partonic distributions, so the two-particle distribution is approximated by a product of single-particle distributions. If the medium contains pair correlations, however, this approximation can miss a distinct contribution to the same microscopic photon-production processes. This raises a more elementary question: how is the local photon emission rate modified when the emitting medium contains pair correlations?

The direct-photon puzzle and the early-time dynamics of relativistic heavy-ion collisions provide the broader phenomenological motivation for this study.

The direct-photon puzzle refers to the difficulty of simultaneously describing the direct-photon yield and anisotropic flow~\cite{PHENIX:2011oxq, Shen:2013vja, PHENIX:2015igl, ALICE:2018dti, David:2019wpt, Monnai:2022hfs, ALICE:2024vwy, Jackson:2025mtk, Kasza:2025wot, Scheid:2025gew, Ayala:2026dgh, Averyanov:2026vvw, Buzzegoli:2026sji, Kroth:2026kgm,Liu:2026dsz}. At low transverse momentum, direct photons exhibit both a large excess yield and a sizable elliptic flow $v_2$. If most photons are emitted at early times, the high temperature can naturally enhance the yield, but the elliptic flow is expected to be small because collective motion has not yet fully developed. If, instead, most photons are emitted at later times, the flow can generate a larger anisotropy, but the lower temperature makes it harder to produce enough photons. This tension suggests that additional microscopic sources, or early-time modifications of the emission rate, may still be missing.

For the latter, the early applicability of hydrodynamics has sharpened the question of what microscopic state is actually required for a fluid-dynamical description~\cite{An:2025ils}. Recent kinetic-theory studies indicate that the onset of hydrodynamic behavior can be separated from the timescale of full thermalization~\cite{Monnai:2023evo, deBrito:2024qow, Du:2025hyk, Lu:2025asx}. This suggests that early photon production should not be viewed only as radiation from a locally equilibrated single-particle distribution, but may also be sensitive to non-equilibrium single-particle distortions and many-body correlations that survive during hydrodynamization.

Since the early medium is a dense many-body system, genuine two-particle correlations provide a natural microscopic structure beyond the factorized rates commonly used in photon phenomenology~\cite{Lu:2025yry}. The spectral BBGKY framework of Ref.~\cite{Lu:2025yry} provides a systematic scheme to incorporate such correlations into kinetic theory. Studying how isotropic pair correlations modify local photon production therefore offers a possible microscopic ingredient for future studies of the direct-photon puzzle, and a complementary probe of correlation dynamics relevant to early hydrodynamization.

The main objective of this paper is to study how correlations in the medium affect the photon production spectrum. As a first step, we focus on the leading-logarithmic contribution from the $2\to2$ quark--antiquark annihilation and Compton channels, rather than the complete leading-order photon emission rate. We write the two-particle distribution as $\mathcal F_{ab}=f_a f_b+g_{ab}$, thereby separating the factorized single-particle contribution from the connected pair correlation. The single-particle distributions and the connected correlation are then expanded in spectral bases, allowing a broad class of two-particle distributions to be represented. In this framework, the microscopic scattering kernels determine how one- and two-particle momentum structures are converted into photon spectral shapes. As an illustrative application, we consider a rotationally invariant medium and show that relative-angle modes of the pair correlation can produce sign-changing modifications to the photon spectrum, with magnitudes comparable to the factorized contribution.

The rest of this paper is organized as follows. In Sec.~\ref{sec:Method}, we formulate the local photon-production rate with a connected two-particle correlation, introduce the spectral representation, and evaluate the response functions for the $2\to2$ annihilation and Compton channels. Then Sec.~\ref{sec:Results} presents the factorized and correlation-induced photon spectra and analyzes how relative-angle correlation modes are mapped into the photon momentum dependence. We summarize, in Sec.~\ref{sec:Conclusion}, the main results and discuss the extensions needed for phenomenological applications. Technical details of the response-function projection and analytic fits to the numerical spectra are given in Appendices~\ref{sec:ResponseFunctions} and~\ref{sec:FitResults}, respectively.

\section{Method}\label{sec:Method}

\subsection{Photon production rate in a local fluid cell}

We consider photon production in a local space-time cell of the medium. The space-time coordinate is denoted by $X$, although it will often be suppressed for notational simplicity. The medium is not assumed to be in local thermal equilibrium, and its microscopic state may contain genuine two-particle correlations.

In this work, we focus on the tree-level $2\to2$ contribution to photon production. For a generic $2\to2$ process,
$
a(p_a)+b(p_b)\rightarrow c(p_c)+\gamma(k),
$
the momentum-differential photon production rate can be written as
\begin{align}
\begin{split}
&
E_k\frac{dR(X)}{d^3k}
=
\sum_{a b\rightarrow c\gamma}
\frac{1}{2(2\pi)^3}
\int
d\Pi_a\, d\Pi_b\, d\Pi_c\,
(2\pi)^4
\\&
\times
\delta^{(4)}(p_a+p_b-p_c-k)
\mathcal K_{ab\rightarrow c\gamma}
\,
\mathcal{S}_{abc}(X;p_a,p_b,p_c).
\end{split}
\end{align}
Here
$
d\Pi_i
=
\frac{d^3p_i}{(2\pi)^3 2E_i}
$
is the Lorentz-invariant phase-space measure. The Dirac-$\delta$ function $\delta^{(4)}(p_a+p_b-p_c-k)$ ensures energy-momentum conservation. $\mathcal K_{ab\rightarrow c\gamma}$ denotes the scattering kernel summed over the intrinsic quantum numbers of both the incoming and outgoing particles. More explicitly, for a given reaction channel $a+b\rightarrow c+\gamma$, we write schematically
$
\mathcal K_{ab\rightarrow c\gamma}
=
\sum_{\text{int.}}
\left|
\sum_{\lambda}
\mathcal M^{(\lambda)}_{ab\rightarrow c\gamma}
\right|^2 .
$
Here $\lambda$ labels the different tree-level diagrams contributing to the same reaction channel. These diagrams are summed at the amplitude level before squaring the amplitude, so that the interference terms are retained. The outer summation $\sum_{\text{int.}}$ denotes the relevant sums over spin (or polarization for massless particles), color, and flavor. This sum includes the intrinsic quantum numbers of the final-state particles as well as those of the incoming particles. It therefore differs from the convention used for $\overline{|\mathcal M|^2}$, where the intrinsic quantum numbers of the incoming particles are averaged rather than summed. Possible degeneracy factors and statistical symmetry factors may also be absorbed into this definition, provided that the same convention is used consistently in the phase-space measure and in the channel sum. The outer sum $\sum_{ab\rightarrow c\gamma}$ in the production rate is different from the diagram sum above. It runs over the distinct photon-production reaction channels, such as Compton-like and annihilation-like processes, with the appropriate particle species assigned to $a$, $b$, and $c$.

The statistical factor entering the photon production rate can be written as
\begin{align}
\mathcal{S}_{abc}(p_a,p_b,p_c)
=
\mathcal{F}_{ab}(p_a,p_b)
+
s_c \mathcal{F}_{abc}^{(3)}(p_a,p_b,p_c),
\end{align}
where $s_c=+1$ for a bosonic final-state medium particle and $s_c=-1$ for a fermionic one. Here
\begin{align}
\mathcal{F}_{ab}(p_a,p_b)
&\equiv
\left\langle
n_a(p_a)n_b(p_b)
\right\rangle ,
\\
\mathcal{F}_{abc}^{(3)}(p_a,p_b,p_c)
&\equiv
\left\langle
n_a(p_a)n_b(p_b)n_c(p_c)
\right\rangle .
\end{align}
We decompose the two-particle distribution as
\begin{align}\label{eq:two_particle_distribution_decomposition}
\mathcal F_{ab}(p_a,p_b)
=
f_a(p_a)f_b(p_b)
+
g_{ab}(p_a,p_b),
\end{align}
where $g_{ab}$ is the connected two-particle correlation.\footnote{This decomposition is fixed by requiring that $g_{ab}$ does not change the single-particle marginals. Defining
\begin{align*}
N_i(X)
\equiv
\int d\Pi_i\, f_i(X;p_i),
\end{align*}
we impose
\begin{align*}
\int d\Pi_a\, \mathcal F_{ab}(X;p_a,p_b)
&=
N_a(X) f_b(X;p_b),
\\
\int d\Pi_b\, \mathcal F_{ab}(X;p_a,p_b)
&=
N_b(X) f_a(X;p_a).
\end{align*}
Equivalently, the connected correlation satisfies the zero-marginal conditions
\begin{align*}
\int d\Pi_a\, g_{ab}(X;p_a,p_b)
=
0,
\qquad
\int d\Pi_b\, g_{ab}(X;p_a,p_b)
=
0.
\end{align*}
With this convention, $g_{ab}$ contains only genuine pair correlations and does not absorb any part of the one-particle distributions. The pair correlation also satisfies the exchange property
\begin{align*}
g_{ab}(X;p_a,p_b)
=
g_{ba}(X;p_b,p_a).
\end{align*}
For identical species, this reduces to the usual symmetry under $p_a\leftrightarrow p_b$.
} 
The three-particle distribution can be decomposed as
\begin{align}
\label{eq:three_particle_distribution_decomposition}
\begin{split}
&
\mathcal F_{abc}^{(3)}(p_a,p_b,p_c)
=
f_a(p_a)f_b(p_b)f_c(p_c)
+
g_{ab}(p_a,p_b)f_c(p_c)
\\
&
+
g_{ac}(p_a,p_c)f_b(p_b)
+
g_{bc}(p_b,p_c)f_a(p_a)
+
h_{abc}(p_a,p_b,p_c),
\end{split}
\end{align}
where $h_{abc}$ denotes the genuine connected three-particle correlation. Substituting these decompositions into $\mathcal S_{abc}$, we obtain
\begin{align}
\begin{split}
\mathcal S_{abc}
=
&
f_a f_b
\left(
1+s_c f_c
\right)
+
g_{ab}
\left(
1+s_c f_c
\right)
\\
&
+
s_c f_b g_{ac}
+
s_c f_a g_{bc}
+
s_c h_{abc}.
\end{split}
\end{align}
For compactness, the momentum arguments have been suppressed in this expression.

Under the correlation decompositions in Eqs.~\eqref{eq:two_particle_distribution_decomposition} and \eqref{eq:three_particle_distribution_decomposition}, the molecular-chaos limit is recovered when all connected correlations vanish,
$
g_{ab}=g_{ac}=g_{bc}=h_{abc}=0.
$
Under this condition, the multiparticle distributions factorize as
$
\mathcal F_{ab}
\rightarrow
f_a f_b$, 
$\mathcal F_{abc}^{(3)}
\rightarrow
f_a f_b f_c ,
$
and the kinetic statistical factor becomes
$
\mathcal S_{abc}
\rightarrow
f_a f_b
\left(
1+s_c f_c
\right).
$

\vspace{3mm}

In this work, we keep two-particle correlations in the medium but neglect genuine three-particle correlations. This corresponds to the closure
\begin{align}
h_{abc}(p_a,p_b,p_c)=0.
\end{align}
The statistical factor then becomes
\begin{align}
\begin{split}
\mathcal S_{abc}^{(2)}
=
&
f_a f_b
\left(
1+s_c f_c
\right)
+
g_{ab}
\left(
1+s_c f_c
\right)
\\
&
+
s_c f_b g_{ac}
+
s_c f_a g_{bc}.
\end{split}
\end{align}
Here, $\mathcal{S}_{abc}^{(2)}$ denotes the closure obtained by retaining two-particle correlations.

Under the two-particle-correlation closure, $h_{abc}=0$, the production rate naturally separates into

\begin{align}
E_k\frac{dR(X)}{d^3k}
=
\left(
E_k\frac{dR(X)}{d^3k}
\right)_{\rm fact.}
+
\left(
E_k\frac{dR(X)}{d^3k}
\right)_{\rm corr.},
\end{align}
where
\begin{align}\label{eq:factor_v1}
\begin{split}
&
\left(
E_k\frac{dR(X)}{d^3k}
\right)_{\rm fact.}
=
\sum_{ab\rightarrow c\gamma}
\frac{1}{2(2\pi)^3}
\int
d\Pi_a\,d\Pi_b\,d\Pi_c\,
(2\pi)^4
\\&\quad\times
\delta^{(4)}(p_a+p_b-p_c-k)
\mathcal K_{ab\rightarrow c\gamma}
f_a f_b
\left(
1+s_c f_c
\right),
\end{split}
\end{align}
and
\begin{align}\label{eq:correlation_v1}
\begin{split}
&
\left(
E_k\frac{dR(X)}{d^3k}
\right)_{\rm corr.}
=
\sum_{ab\rightarrow c\gamma}
\frac{1}{2(2\pi)^3}
\int
d\Pi_a\,d\Pi_b\,d\Pi_c\,
(2\pi)^4
\\&\quad\times
\delta^{(4)}(p_a+p_b-p_c-k)
\mathcal K_{ab\rightarrow c\gamma}
\\&\quad\times
\left[
g_{ab}
\left(
1+s_c f_c
\right)
+
s_c f_b g_{ac}
+
s_c f_a g_{bc}
\right].
\end{split}
\end{align}
In these expressions, $f_i=f_i(X;p_i)$, $g_{ij}=g_{ij}(X;p_i,p_j)$.

\subsection{Isotropic off-equilibrium medium and angular pair correlations}

To keep the structure of the photon emission rate transparent, we assume that all medium components share the same non-equilibrium deformation and the same correlation profile, while their Bose or Fermi statistical factors remain species dependent. More explicitly, for a medium particle of species $a$, we write
\begin{align}
f_a(X;p)
=
f_a^{\rm ref}(X;p)\,\Phi(X;p),
\end{align}
where
\begin{align}
f_a^{\rm ref}(X;p)
=
\frac{1}{\exp\!\left(\frac{p \cdot u(X)}{\Lambda(X)}\right)+s_a}.
\end{align}
Here $u^\mu(X)$ denotes the local rest-frame four-velocity used to define the energy variable $p\cdot u(X)$, and $\Lambda(X)$ is a local reference momentum scale. It need not be a thermodynamic temperature, since the fluid cell is not assumed to be in local thermal equilibrium. In the equilibrium limit, $\Lambda(X)$ can be identified with the local temperature. The parameter $s_a=-1$ for gluons and $s_a=+1$ for quarks and antiquarks. The function $\Phi(X;p)$ describes a species-independent non-equilibrium deformation relative to this reference distribution. The reference equilibrium form is recovered for $\Phi(X;p)=1$.

The same convention is used for the pair correlation. The connected correlation contribution to the two-particle distribution $g_{ab}(X;p_1,p_2)$ is parameterized as
\begin{align}
g_{ab}(X;p_1,p_2)
=
f_a^{\rm ref}(X;p_1)f_b^{\rm ref}(X;p_2)\,
\Psi(X;p_1,p_2).
\end{align}
Here $\Psi(p_1,p_2)$ is taken to be a species independent correlation profile, while the quantum statistical factors remain species dependent through $f_a^{\rm ref}$ and $f_b^{\rm ref}$.

\vspace{3mm}

In this work, we restrict the off-equilibrium medium to isotropic deviations in the local rest frame. Thus, the one-particle distribution depends only on the local-rest-frame energy variable
$
\xi\equiv \frac{p\cdot u(X)}{\Lambda(X)},
$
and is parameterized as
\begin{align}
f_a(X;p)=f_a^{\rm ref}(X;\xi)\,\Phi(X;\xi).
\end{align}

For the correlated two-particle distribution, rotational invariance allows the connected correlation to depend on the two local-rest-frame energies and on the relative angle between the two momenta. Thus,
\begin{align}
g_{ab}(X;p_1,p_2)
=
f_a^{\rm ref}(X;\xi_1)f_b^{\rm ref}(X;\xi_2)\,
\Psi(X;\xi_1,\xi_2,z_{12}),
\end{align}
where
\begin{align}
\xi_i
\equiv
\frac{p_i\cdot u(X)}{\Lambda(X)}
\quad
z_{12}
\equiv
\hat{\bp}_1\cdot \hat{\bp}_2 .
\end{align}
Here $\hat{\bp}_i$ denotes the direction of the three-momentum of particle $i$ in the local rest frame. The variable $z_{12}$ is invariant under a common rotation of $\bp_1$ and $\bp_2$ in that frame. Thus, even in an isotropic medium, the pair correlation may contain nontrivial angular dependence. This angular dependence does not introduce a preferred direction for the medium. Instead, it describes how two particles are correlated with each other inside an otherwise isotropic system.

The explicit argument $X$ in $\Phi(X;\xi)$ and $\Psi(X;\xi_1,\xi_2,z_{12})$ denotes possible local space-time dependence beyond the dependence already contained in the dimensionless variables through $u^\mu(X)$ and the local reference momentum scale $\Lambda(X)$. More general local ansatzes may be written as
\begin{align}
\Phi(X;p)
&=
\Phi\!\left(
\xi;
\mathcal Q_\Phi(X)
\right),
\\
\Psi(X;p_1,p_2)
&=
\Psi\!\left(
\xi_1,\xi_2,z_{12};
\mathcal Q_\Psi(X)
\right),
\end{align}
where
\begin{align}
\mathcal Q_\Phi(X)
&=
\left\{
n,\mu,\nabla_\mu \Lambda,\nabla_\mu u_\nu,
\mathcal I_\Phi
\right\},
\\
\mathcal Q_\Psi
&=
\left\{
n,\mu,\nabla_\mu \Lambda,\nabla_\mu u_\nu,
\ell_{\rm corr},\mathcal I_\Psi
\right\}.
\end{align}
The sets $\mathcal Q_\Phi(X)$ and $\mathcal Q_\Psi(X)$ represent additional local information that is not fixed only by $u^\mu(X)$ and $\Lambda(X)$. They may include the local density $n(X)$, the chemical potential $\mu(X)$, gradients of the local reference scale and flow field, such as $\nabla_\mu \Lambda(X)$ and $\nabla_\mu u_\nu(X)$, and residual nonhydrodynamic information. The correlation profile $\Psi$ may also depend on a local correlation length $\ell_{\rm corr}(X)$.
It characterizes the typical spatial range over which two-particle correlations remain important in the medium. In a near-equilibrium system, this scale may be related to microscopic interaction lengths or relaxation scales, while in a strongly non-equilibrium system it may also reflect the size of correlated domains generated during the earlier evolution.
The quantities $\mathcal I_\Phi(X)$ and $\mathcal I_\Psi(X)$ denote residual nonhydrodynamic information inherited from the initial state or from the previous dynamical history of the system. They represent memory effects that are not fully specified by the local fields $u^\mu(X)$, $\Lambda(X)$, or by their gradients. Examples include remnants of initial-state fluctuations, local parton production patterns, or other slowly relaxing non-equilibrium structures.

Under the assumptions that all particle species share the same distribution function and that the system is rotationally invariant, the factorized contribution in Eq.~\eqref{eq:factor_v1} takes the form
\begin{align}\label{eq:factor_v2}
\begin{split}
&
\left(
E_k\frac{dR(X)}{d^3k}
\right)_{\rm fact.}
=
\sum_{ab\rightarrow c\gamma}
\frac{1}{2(2\pi)^3}
\int
d\Pi_a\,d\Pi_b\,d\Pi_c\,
(2\pi)^4
\\&\times
\delta^{(4)}(p_a+p_b-p_c-k)
\mathcal K_{ab\rightarrow c\gamma}
f(\xi_a)f(\xi_b)
\big(1+s_c f(\xi_c)\big),
\end{split}
\end{align}
while the correlation contribution in Eq.~\eqref{eq:correlation_v1} becomes
\begin{align}\label{eq:correlation_v2}
\begin{split}
&
\left(
E_k\frac{dR(X)}{d^3k}
\right)_{\rm corr.}
=
\pi\sum_{ab\rightarrow c\gamma}
\int d\Pi_a\,d\Pi_b\,d\Pi_c\,
\mathcal{K}_{ab\rightarrow c\gamma}
\\&\;
\delta^{(4)}(p_a+p_b-p_c-k)
\;
\Big(
g(\xi_a,\xi_b,z_{ab})
\big(1+s_c f(\xi_c)\big)
\\&
+ s_c f(\xi_b)g(\xi_a,\xi_c,z_{ac})
+ s_c f(\xi_a)g(\xi_b,\xi_c,z_{bc})
\Big).
\end{split}
\end{align}

After the phase-space integration, rotational invariance implies that the photon momentum can enter the rate only through $|\bk| \equiv k\cdot u(X)$. We define
\begin{align}
\frac{dR^{\rm fac}(X)}{d^3k}
&\equiv
\mathcal R^{\rm fac}(X;|\bk|),
\\
\frac{dR^{\rm corr}(X)}{d^3k}
&\equiv
\mathcal R^{\rm corr}(X;|\bk|).
\end{align}
Therefore,
$
E_k\frac{dR(X)}{d^3k}
=
E_k\mathcal R^{\rm fac}(X;|\bk|)
+
E_k\mathcal R^{\rm corr}(X;|\bk|).
$

\subsection{Spectral Representation of Isotropic Non-Equilibrium Distributions}

In a QGP produced in heavy-ion collisions, non-equilibrium distributions and correlations can depend on the initial state, the expansion history, and the microscopic scattering dynamics. To obtain a systematic and computable expression for the photon production rate, we represent their momentum dependence by a spectral expansion.

We expand the single-particle non-equilibrium deformation $\Phi(X,\xi)$ in generalized Laguerre polynomials,
\begin{align}
\Phi(X,\xi)
=
\sum_{n=0}^{N}
\Phi_n(X) L_n^{(2)}(\xi).
\end{align}
The upper limit $N$ is shown explicitly to indicate the practical truncation; it will be omitted in later expressions for notational simplicity. The choice of $L_n^{(2)}$ is natural for isotropic momentum integrals. These polynomials are orthogonal on the semi-infinite domain $\xi\in[0,\infty)$ with the weight $\xi^2 e^{-\xi}$,
$
\int_0^\infty d\xi\, \xi^2 e^{-\xi}
L_n^{(2)}(\xi)L_m^{(2)}(\xi)
=
\frac{(n+2)!}{n!}\delta_{nm}.
$
The factor $\xi^2$ comes from the radial phase-space measure, while $e^{-\xi}$ represents the typical exponential falloff of thermal-like distributions.

For the connected two-particle correlation, we use a product basis in the two radial momenta and a Legendre expansion in the relative angle,
\begin{align}
\begin{split}
&
    \Psi(X;\xi_1,\xi_2,z_{12})
\\=\;&
    \sum_{n,m=0}^{N}
    \sum_{\ell=0}^{L}
    \Psi_{nm\ell}(X)\,
    L_n^{(2)}(\xi_1)L_m^{(2)}(\xi_2)
    P_\ell(z_{12}).
\end{split}
\end{align}
The Legendre index $\ell$ describes the relative-angle structure of the pair correlation. It does not introduce a preferred direction for the medium.

The expansion coefficients $\Phi_n$ and $\Psi_{nm\ell}$ encode the medium information. They are not arbitrary if the distribution is required to be physical. After truncation, the coefficients should be chosen such that
\begin{align}
f_i(X,p)&\ge 0,
\\
f_a(X,p_a)f_b(X,p_b)+g_{ab}(X;p_a,p_b,z_{ab})&\ge 0,
\end{align}
in the momentum region relevant to the calculation. 

After inserting the spectral expansions, for a given reaction channel $a+b \rightarrow c+\gamma$ labeled by $\chi$, the factorized contribution is
\begin{align}\label{eq:factor_v3}
\mathcal R_{\chi}^{\rm fac}
=
\sum_{n,m}
\Phi_{n}\Phi_{m}
\mathcal I^{ab;\chi,0}_{nm}
+
s_c
\sum_{n,m,r}
\Phi_{n}\Phi_{m}\Phi_{r}
\mathcal I^{abc;\chi,1}_{nmr}.
\end{align}
The first term comes from the product $f_a f_b$, while the second term arises from the final-state statistical factor and is proportional to $f_a f_b f_c$. The correlation-induced contribution is
\begin{align}\label{eq:correlation_v3}
\begin{split}
&
\mathcal R_{\chi}^{\rm corr}
=\;
\sum_{n,m,\ell}
\Psi_{nm\ell}
\mathcal J^{ab;\chi,0}_{nm\ell}
+
s_c
\sum_{n,m,\ell,r}
\Psi_{nm\ell}\Phi_{r}
\mathcal J^{ab,c;\chi,1}_{nm\ell r}
\\&
+
s_c
\sum_{n,m,\ell,r}
\Psi_{nm\ell}\Phi_{r}
\mathcal J^{ac,b;\chi,1}_{nm\ell r}
+
s_c
\sum_{n,m,\ell,r}
\Psi_{nm\ell}\Phi_{r}
\mathcal J^{bc,a;\chi,1}_{nm\ell r}.
\end{split}
\end{align}
The first term is the direct contribution from the initial-state pair correlation $g_{ab}$, while the remaining terms arise from the final-state statistical factor combined with pair correlations involving $c$.

The basis-response functions are defined by
\begin{align}
\label{eq:response-I-ab-0}
\begin{split}
&
\mathcal{I}^{ab;\chi,0}_{nm}
=\;
\frac{1}{2(2\pi)^3 E_k}
\int d\Pi_a d\Pi_b d\Pi_c\,
\mathcal K_\chi\,(2\pi)^4
\\&\times
\delta^{(4)}(p_a+p_b-p_c-k)
\,\mathcal B_{a,n}(\xi_a)\mathcal B_{b,m}(\xi_b),
\end{split}
\\
\label{eq:response-I-abc-1}
\begin{split}
&
\mathcal{I}^{abc;\chi,1}_{nmr}
=\;
\frac{1}{2(2\pi)^3 E_k}
\int d\Pi_a d\Pi_b d\Pi_c\,
\mathcal K_\chi\,(2\pi)^4
\\&\times
\delta^{(4)}(p_a+p_b-p_c-k)
\mathcal B_{a,n}(\xi_a)\mathcal B_{b,m}(\xi_b)\mathcal B_{c,r}(\xi_c),
\end{split}
\\
\label{eq:response-J-ab-0}
\begin{split}
&
    \mathcal J^{ab;\chi,0}_{nm\ell}
=\;
\frac{1}{2(2\pi)^3 E_k}
\int d\Pi_a d\Pi_b d\Pi_c\,
\mathcal K_\chi\,(2\pi)^4
\\&\times
\delta^{(4)}(p_a+p_b-p_c-k)
\mathcal B_{a,n}(\xi_a)\mathcal B_{b,m}(\xi_b)P_\ell(z_{ab}),
\end{split}
\\
\label{eq:response-J-abc-1}
\begin{split}
&
    \mathcal J^{ab,c;\chi,1}_{nm\ell r}
=\;
\frac{1}{2(2\pi)^3 E_k}
\int d\Pi_a d\Pi_b d\Pi_c\,
\mathcal K_\chi\,(2\pi)^4
\\&\times
\delta^{(4)}(p_a+p_b-p_c-k)
\mathcal B_{a,n}(\xi_a)\mathcal B_{b,m}(\xi_b)\mathcal B_{c,r}(\xi_c)
P_\ell(z_{ab}),
\end{split}
\end{align}
where
$
\mathcal B_{a,n}(\xi)\equiv f_a^{\rm ref}(\xi) L_n^{(2)}(\xi).
$
The functions $\mathcal J^{ac,b;\chi,1}$ and $\mathcal J^{bc,a;\chi,1}$ are obtained by the corresponding permutations of particle labels.

These response functions depend only on the reaction channel, the phase-space constraint, and the chosen spectral basis. Once they are computed, different non-equilibrium media can be studied by changing only the coefficients $\Phi_n$ and $\Psi_{nm\ell}$. The continuous functional problem of evaluating the photon rate is therefore reduced to a finite set of basis-response functions and a finite set of medium coefficients.

The detailed reduction of the phase-space integrals is algebraic and not essential to the physical argument in the main text. We therefore keep the response functions in the compact form introduced above. Their explicit derivation and computational details are given in Appendix~\ref{sec:ResponseFunctions}.

\subsection{Calculation for the $2\to2$ channels}

We now apply the above framework to the leading $2\to2$ photon-production channels in a QGP.

The medium is composed of quarks, antiquarks, and gluons. In the temperature range relevant for the applications considered here, with a typical momentum scale of a few hundred MeV, we include the active light flavors $f=u,d,s$ as thermal medium constituents and do not include heavier quarks. The light quarks are treated as massless because their masses are small compared with the typical momentum scale of the medium, while gluons are massless gauge bosons. For all partons considered here, this gives \(p_i\cdot u(X)=|\bp_i|\) in the local rest frame.

\begin{figure}
\centering
\includegraphics[width=1\linewidth]{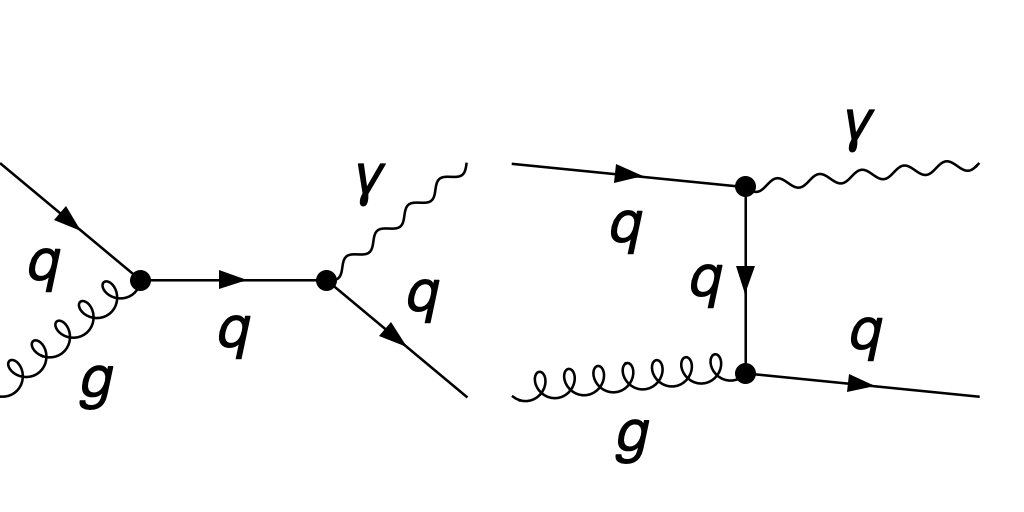}
\caption{Compton scattering process.}
\label{fig:Compton}
\end{figure}

\begin{figure}
\centering
\includegraphics[width=1\linewidth]{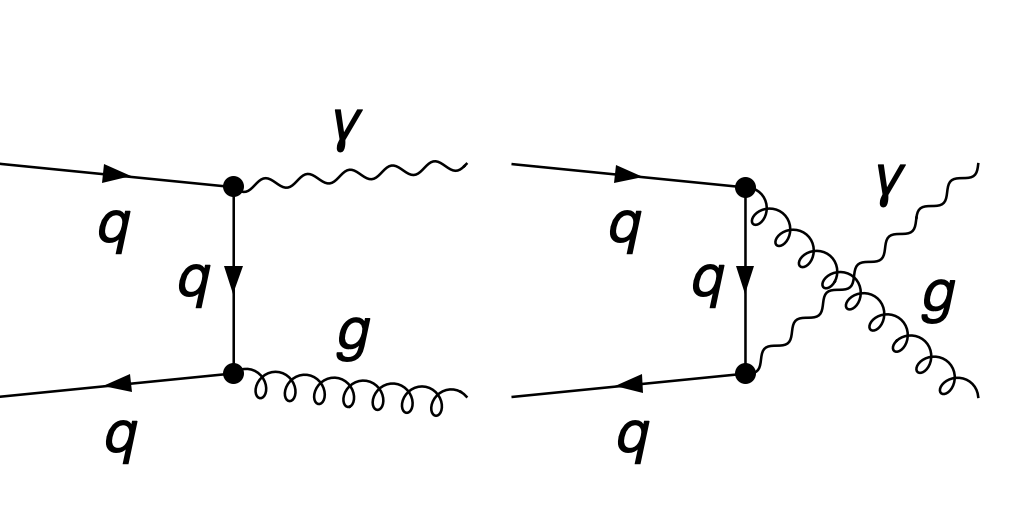}
\caption{Annihilation process.}
\label{fig:Annihilation}
\end{figure}

We consider the Compton scattering processes,
\begin{align}
q_f+g\rightarrow q_f+\gamma,
\qquad
\bar q_f+g\rightarrow \bar q_f+\gamma,
\end{align}
and the annihilation process,
\begin{align}
q_f+\bar q_f\rightarrow g+\gamma.
\end{align}
Here $f$ denotes the quark flavor. The corresponding diagrams are shown in Figs.~\ref{fig:Compton} and \ref{fig:Annihilation}.


The quantity that enters the phase-space integrals in Eq.~\ref{eq:response-I-ab-0}--\ref{eq:response-J-abc-1} is the initial-state summed kernel $\mathcal K_\chi$. In the following, however, we first quote the squared matrix element in the common quantum-field-theory convention, where it is averaged over the spin and color states of the incoming particles and summed over those of the outgoing particles. We denote this initial-state averaged quantity by $\overline{|\mathcal M_\chi|^2}$. The kernel $\mathcal K_\chi$ used in the rate is then obtained by multiplying $\overline{|\mathcal M_\chi|^2}$ by the spin-color degeneracy of the incoming medium particles.

To write the channel kernels below, we use a common convention for the Mandelstam variables. For a channel labeled as $a(p_a)+b(p_b)\rightarrow c(p_c)+\gamma(k)$, we define
$
s=(p_a+p_b)^2,
t=(p_a-k)^2,
u=(p_b-k)^2 .
$
Since the external particles are treated as massless in this work, these variables satisfy
$
s+t+u=0 .
$

For the annihilation channel,
$
q_f(p_a)+\bar q_f(p_b)\rightarrow g(p_c)+\gamma(k).
$
For a fixed flavor $f$, the initial-state averaged squared matrix element in the convention specified above is
$
\overline{\left|\mathcal{M}_{\rm ann}^{(f)}\right|^2}
=
\frac{N_c^2-1}{N_c^2}\,
e^2 g_s^2 Q_f^2
\left(
\frac{u}{t}
+
\frac{t}{u}
\right).
$
The factor $(N_c^2-1)/N_c^2$ is the color factor of the annihilation process. The numerator $N_c^2-1$ counts the number of color states of the final-state gluon, which transforms in the adjoint representation of $SU(N_c)$. The denominator $N_c^2$ counts the number of color states of the incoming quark-antiquark pair, with each fermion transforming in the fundamental representation of $SU(N_c)$.~\footnote{Here, ``transforms'' is used in the standard group-theoretical sense: the color degrees of freedom of the particle are acted on by an $SU(N_c)$ gauge transformation according to a specified representation.} For QCD, $N_c=3$, and the color factor becomes $8/9$. The factor $e$ is the electromagnetic gauge coupling, equivalently the magnitude of the elementary electric charge, and $g_s$ is the QCD strong gauge coupling. The flavor-summed initial-state averaged squared matrix element $\overline{|\mathcal{M}_{\rm ann}|^2}=\sum_{f=u,d,s} \overline{\left|\mathcal{M}_{\rm ann}^{(f)}\right|^2}$ is therefore
\begin{align}\label{eq:M2_ann_flavor_sum}
\begin{split}
\overline{|\mathcal{M}_{\rm ann}|^2}
&=
\frac{N_c^2-1}{N_c^2}\,
e^2 g_s^2
\left(
\sum_{f=u,d,s} Q_f^2
\right)
\left(
\frac{u}{t}
+
\frac{t}{u}
\right),
\end{split}
\end{align}
where
$
    \sum_{f=u,d,s} Q_f^2
=
    Q_u^2+Q_d^2+Q_s^2
=
\left(\frac{2}{3}\right)^2
+
\left(-\frac{1}{3}\right)^2
+
\left(-\frac{1}{3}\right)^2
=
\frac{2}{3}
$.
To obtain the kernel summed over the spin and color states of the incoming quark and antiquark, one multiplies the averaged result by $(2N_c)(2N_c)=4N_c^2$. Thus\footnote{For the annihilation channel, this normalization is equivalent to the AMY normalization after using the symmetry of the phase-space integral. AMY writes the annihilation kernel as $16\,e^2 g_s^2(\sum_f Q_f^2)d_F C_F\, (u_{\rm AMY}/t_{\rm AMY})$, with $d_F=N_c$ and $C_F=(N_c^2-1)/(2N_c)$, so $16d_FC_F=8(N_c^2-1)$. The AMY variables $t_{\rm AMY}$ and $u_{\rm AMY}$ are tied to their ordering of the incoming quark and antiquark. Our initial-state summed annihilation kernel contains $4(N_c^2-1)(u/t+t/u)$. In a charge-symmetric medium, the annihilation phase-space integral is invariant under $p_q\leftrightarrow p_{\bar q}$, which exchanges $t\leftrightarrow u$. Therefore $\int(t/u)=\int(u/t)$, and our annihilation kernel is equivalent under the integral to $8(N_c^2-1)t/u$. Since the AMY $t_{\rm AMY},u_{\rm AMY}$ labels are interchanged relative to the present $t,u$ labels, $u_{\rm AMY}/t_{\rm AMY}=t/u$, and the two conventions agree.}
\begin{align}\label{eq:K_ann_flavor_sum}
\begin{split}
\mathcal K_{\rm ann}
&=
4N_c^2\,
\overline{|\mathcal{M}_{\rm ann}|^2}
\\
&=
4\left(N_c^2-1\right)
e^2 g_s^2
\left(
\sum_{f=u,d,s} Q_f^2
\right)
\left(
\frac{u}{t}
+
\frac{t}{u}
\right).
\end{split}
\end{align}

Eq.~\eqref{eq:M2_ann_flavor_sum} can also be written in terms of the electromagnetic coupling $\alpha_{\rm EM}=e^2/(4\pi)$ and the strong coupling $\alpha_s=g_s^2/(4\pi)$. In this form, the result is proportional to $\alpha_{\rm EM}\alpha_s$. This factor reflects the leading-order (tree-level) structure of the process. The amplitude for each flavor contains one photon--quark vertex, proportional to $Q_f e$, and one quark--gluon vertex, proportional to $g_s$, so the squared amplitude is of order $e^2 g_s^2$, or equivalently $O(\alpha_{\rm EM}\alpha_s)$. 

For the Compton channel, $q_f(p_a)+g(p_b)\rightarrow q_f(p_c)+\gamma(k)$, the flavor-summed initial-state averaged squared matrix element $\overline{|\mathcal{M}_{\rm Comp}|^2}=\sum_{f=u,d,s} \overline{\left|\mathcal{M}_{\rm Comp}^{(f)}\right|^2}$ is
\begin{align}\label{eq:M2_Comp_flavor_sum}
\begin{split}
\overline{|\mathcal{M}_{\rm Comp}|^2}
&=
\frac{1}{N_c}\,
e^2 g_s^2
\left(
\sum_{f=u,d,s} Q_f^2
\right)
\left(
-
\frac{s}{u}
-
\frac{u}{s}
\right).
\end{split}
\end{align}
The kernel entering the rate integral is obtained by undoing the initial-state spin-color average. For the quark Compton channel, this amounts to multiplying by the incoming quark and gluon degeneracy, $(2N_c)\,[2(N_c^2-1)]=4N_c(N_c^2-1)$. Thus
\begin{align}\label{eq:K_Comp_flavor_sum}
\begin{split}
\mathcal K_{\rm Comp}^{q}
&=
4N_c(N_c^2-1)\,
\overline{|\mathcal{M}_{\rm Comp}|^2}
\\
&=
4\left(N_c^2-1\right)
e^2 g_s^2
\left(
\sum_{f=u,d,s} Q_f^2
\right)
\left(-
\frac{s}{u}
-
\frac{u}{s}
\right).
\end{split}
\end{align}
The antiquark Compton channel, $\bar q_f(p_a)+g(p_b)\rightarrow \bar q_f(p_c)+\gamma(k)$, has the same squared matrix element, with the same charge factor $Q_f^2$.\footnote{For the Compton channel, the quark and antiquark Compton contributions are written as separate channels here. In a charge-symmetric medium they are equal, so their sum gives an extra factor of two: $\mathcal K_{\rm Comp}^{q}+\mathcal K_{\rm Comp}^{\bar q}=8(N_c^2-1)e^2g_s^2(\sum_f Q_f^2)(-s/u-u/s)$. AMY writes the Compton kernel as $16\,e^2 g_s^2(\sum_f Q_f^2)d_F C_F(-s/t_{\rm AMY}-t_{\rm AMY}/s)$. Since $16d_FC_F=8(N_c^2-1)$ and $t_{\rm AMY}=u$ in the convention used here, this becomes $8(N_c^2-1)e^2g_s^2(\sum_f Q_f^2)(-s/u-u/s)$, matching the present Compton kernel after summing the quark and antiquark channels.}

Using these channel-dependent initial-state summed kernels, the total leading $2\to2$ photon-production rate is obtained by summing over the annihilation, quark Compton, and antiquark Compton channels.

The full leading $2\to2$ photon-production rate is therefore
\begin{align}
\begin{split}
&
    E_k\frac{dR^{2\to2}(X)}{d^3k}
    =
    E_k\frac{dR_{\rm ann}(X)}{d^3k}
\\&\qquad\qquad +\;
    E_k\frac{dR_{\rm Comp}^{q}(X)}{d^3k}
    +
    E_k\frac{dR_{\rm Comp}^{\bar q}(X)}{d^3k}.
\end{split}
\label{eq:total_2to2_photon_rate}
\end{align}
Equivalently,
$
    \mathcal R^{2\to2}(X;k)
=
    \mathcal R_{\rm ann}(X;k)
    +
    \mathcal R_{\rm Comp}^{q}(X;k)
    +
    \mathcal R_{\rm Comp}^{\bar q}(X;k),
$
where
$
\mathcal R_\chi(X;k) \equiv \frac{dR_\chi(X)}{d^3k}.
$
If the medium is charge symmetric, so that the quark and antiquark distributions and correlations are identical, the two Compton contributions are equal. In that case,
$
E_k\frac{dR^{2\to2}(X)}{d^3k}
=
E_k\frac{dR_{\rm ann}(X)}{d^3k}
+
2\,E_k\frac{dR_{\rm Comp}^{q}(X)}{d^3k}.
\label{eq:total_2to2_charge_symmetric}
$

\section{Results}\label{sec:Results}

In this section, we exploit a few examples to show, explicitly, how two-particle correlations modify the photon spectrum.

For clarity, we work in the classical-statistics limit. More explicitly, we neglect Bose enhancement and Pauli blocking factors and replace the equilibrium Bose-Einstein and Fermi-Dirac distribution functions by their Boltzmann limits. The single-particle distribution is then written as
\begin{align}
f(X,p)
=
e^{-\xi}\Phi(X,\xi).
\end{align}
The connected two-particle correlation is parameterized in an analogous form,
\begin{align}
g_{ab}(X;p_1,p_2)
=
e^{-\xi_1-\xi_2}
\Psi(X;\xi_1,\xi_2,z_{12}).
\end{align}
This approximation keeps the leading Boltzmann term in the fugacity expansion and provides a clean baseline for isolating the effect of pair correlations before restoring quantum-statistical factors. In this limit, all terms in Eqs.~\eqref{eq:factor_v3} and \eqref{eq:correlation_v3} that are proportional to $s_c$ vanish.

\begin{figure}
\centering
\includegraphics[width=1\linewidth]{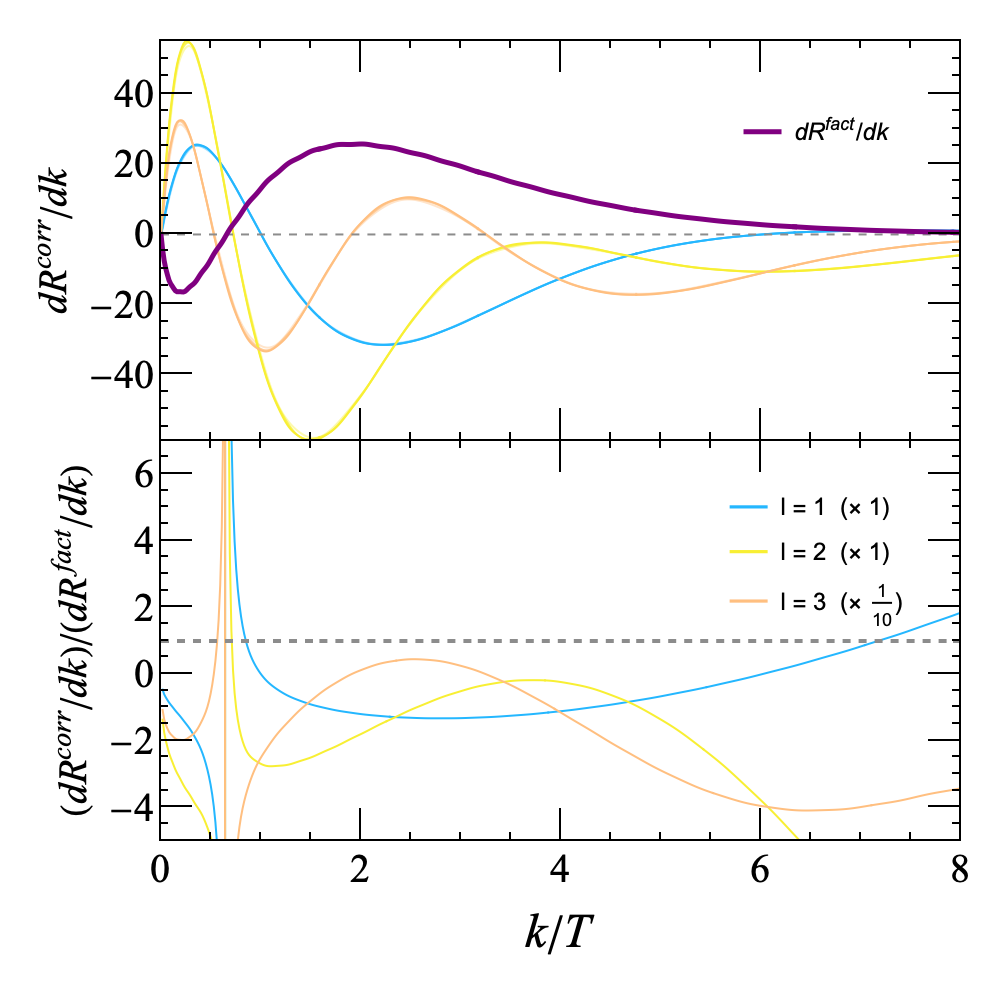}
\caption{Leading-logarithmic photon spectrum from the quark--antiquark annihilation channel in the classical-statistics limit. For the factorized baseline, we set the single-particle deformation to $\Phi(\xi)=1$; the corresponding contribution $dR^{\rm fact}/dk$ is shown by the thick purple curve. The thin colored curves show the correlation-induced contributions $dR^{\rm corr}/dk$ generated by the lowest radial correlation mode, $n=m=0$, for relative-angle modes $\ell=1,2,3$. The rescaling factors used for visual clarity are indicated in the legend. The lower panel shows the ratio of each correlation-induced contribution to the factorized contribution.}
\label{fig:AnnihilationResult}
\end{figure}

To isolate the effect of two-particle correlations on the photon spectrum, we set the single-particle off-equilibrium deformation to unity, $\Phi(\xi)=1$. In this limit, the one-particle distribution functions reduce to their equilibrium forms. Therefore, any deviation from the factorized baseline can be attributed to the connected two-particle sector. The corresponding annihilation leading-logarithmic contribution without pair correlations is shown by the thick purple curve in the upper panel of Fig.~\ref{fig:AnnihilationResult}. 

The negative values at low momentum are not physical rates; they indicate that the leading-logarithmic, classical-statistics approximation is being applied outside the region where it is reliable. In particular, the nonlogarithmic pieces of the full HTL-resummed leading-order $2\to2$ rate, as well as quantum-statistical factors, are not included in this curve.

We then turn on a controlled set of isotropic two-particle correlations. For illustration, we consider the lowest radial sector of the correlation basis, $n=m=0$, and vary only the relative-angle mode $\ell$. The $\ell=0$ component is not shown in this comparison, because in this simple limit it only produces a momentum-independent rescaling of the factorized contribution,
$
\mathcal F_{ab}
=
f_a f_b + g_{ab}
\propto
f_a f_b .
$
This component therefore only rescales the factorized contribution in the present setup. The colored thin curves in the upper panel of Fig.~\ref{fig:AnnihilationResult} show the resulting correlation-induced contributions for $\ell\neq 0$. For visual clarity, these curves have been rescaled by the factors indicated in the legend.

The lower panel of Fig.~\ref{fig:AnnihilationResult} shows the ratio of each correlation-induced contribution to the factorized equilibrium contribution. The ratio alternates between enhancement and suppression as a function of $k/T$. This demonstrates that the relative-angle structure of the pair correlation is converted by the annihilation kernel into a nontrivial energy dependence of the emitted photon spectrum. Thus, even when the medium is rotationally invariant and the single-particle distributions are thermal, two-particle correlations can modify the photon spectral shape.

\begin{figure}
\centering
\includegraphics[width=1\linewidth]{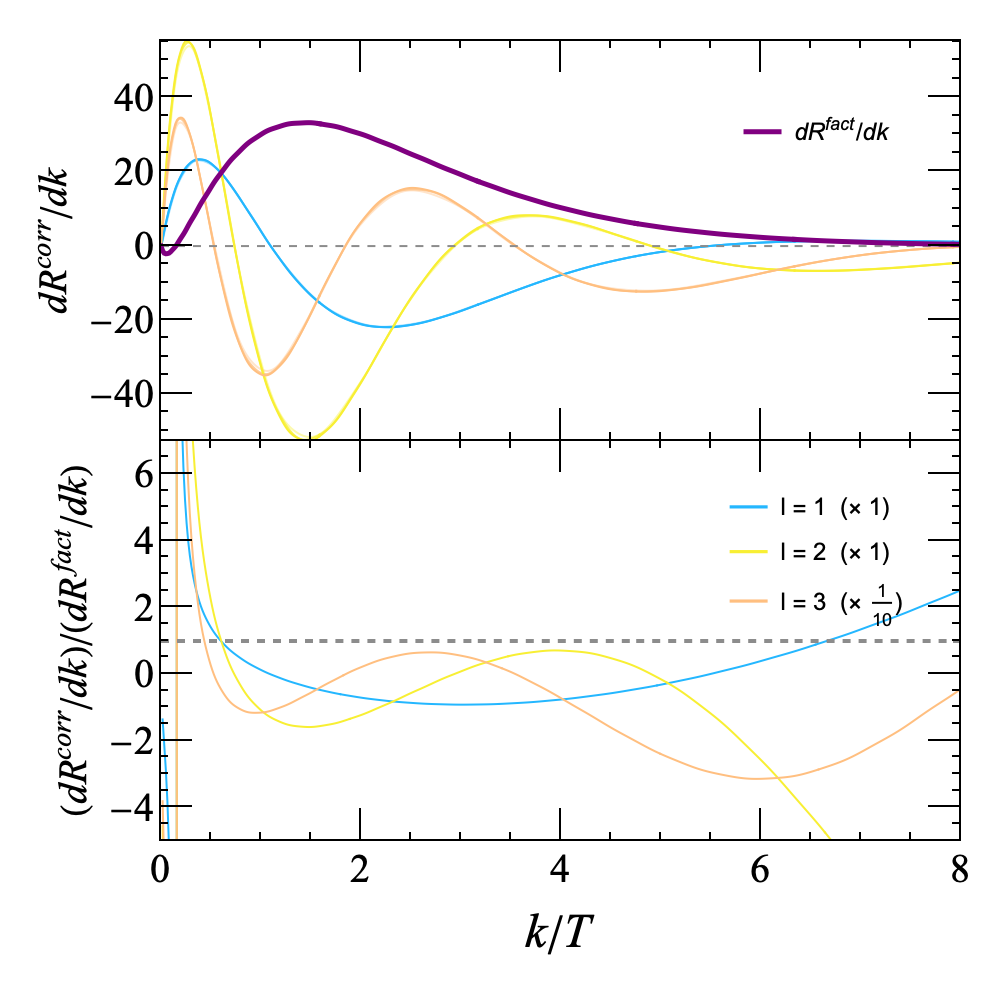}
\caption{Same as Fig.~\ref{fig:AnnihilationResult}, but for the Compton channel. The thick purple curve is the factorized contribution, while the thin colored curves show the correlation-induced contributions from the $n=m=0$, $\ell=1,2,3$ pair-correlation modes. 
The lower panel gives the corresponding ratios to the factorized contribution.}
\label{fig:ComptonResult}
\end{figure}

The corresponding results for the Compton process are shown in Fig.~\ref{fig:ComptonResult}. As in the annihilation channel, the factorized contribution gives a smooth spectrum. The negative low-momentum region should again be interpreted as a limitation of the leading-logarithmic, classical-statistics approximation rather than as a physical photon-production rate. By contrast, the nonzero-$\ell$ correlation components generate sign-changing corrections in momentum space. This shows that the Compton kernel also maps rotationally invariant two-particle correlations into nontrivial modifications of the photon energy spectrum. The detailed momentum dependence differs from the annihilation case, reflecting the different structure of the underlying scattering kernel.

\begin{figure}
\centering
\includegraphics[width=1\linewidth]{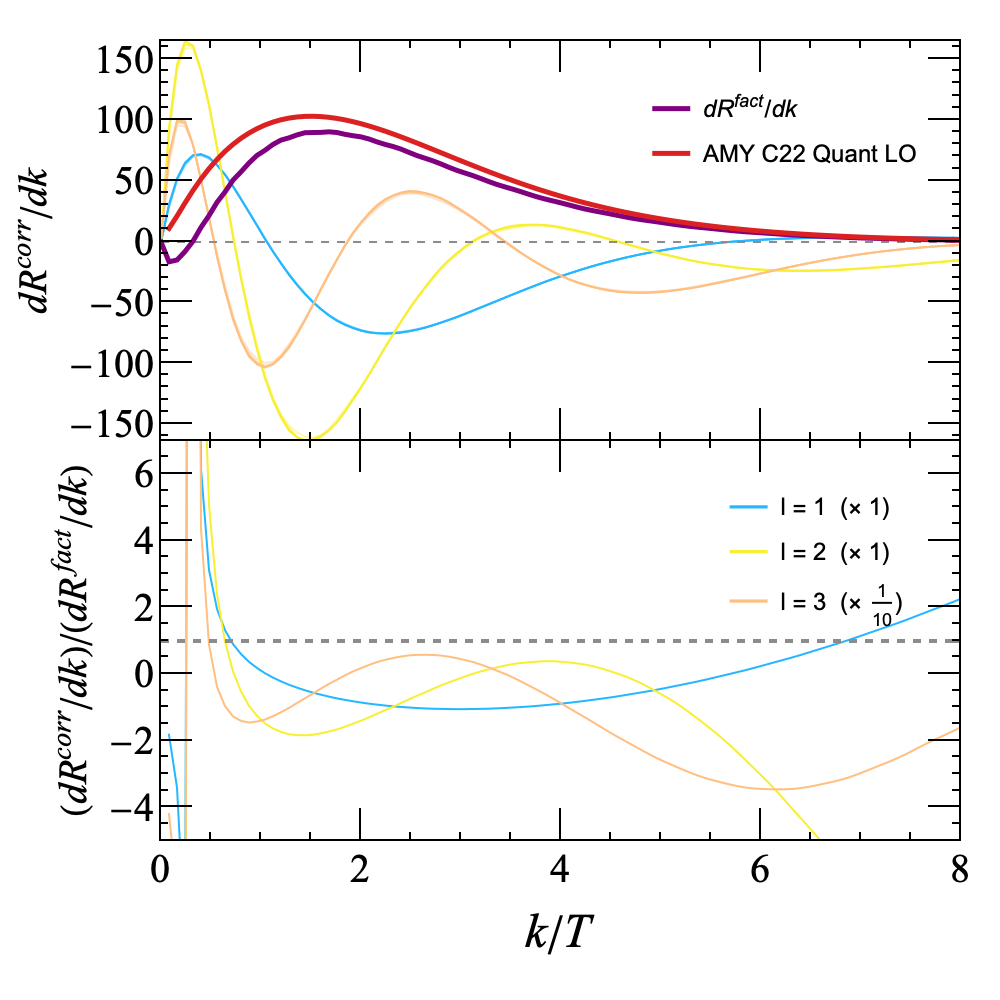}
\caption{Total $2\to2$ photon spectrum from the annihilation and Compton channels. The factorized leading-logarithmic result is compared with the AMY $2\to2$ benchmark, while the colored curves show correlation-induced contributions from $n=m=0$, $\ell=1,2,3$ pair-correlation modes. The lower panel shows their ratios to the factorized contribution.}
\label{fig:PhotonResult}
\end{figure}

The total photon-production spectrum, including the annihilation, quark Compton, and antiquark Compton channels, is shown in Fig.~\ref{fig:PhotonResult}. The red curve shows the $2\to2$ contribution to the leading-order thermal photon-production rate of Arnold, Moore, and Yaffe (AMY)~\cite{Arnold:2001ms}, including the quantum-statistical distribution functions. It is included as a benchmark for the leading-logarithmic $2\to2$ calculation used here.

Since the AMY result is a full leading-order calculation, while the present result keeps only the leading-logarithmic $2\to2$ contribution in the classical-statistics limit, quantitative agreement is not expected over the whole momentum range. The comparison improves at large $k/T$, where the equilibrium quantum distribution functions approach their Boltzmann limits. The remaining difference should be attributed to the nonlogarithmic leading-order contributions.

The correlation-induced contributions in the total spectrum display a clear oscillatory behavior as functions of $k/T$. As in Figs.~\ref{fig:AnnihilationResult} and \ref{fig:ComptonResult}, they change sign several times and decrease gradually at large momentum. This behavior reflects the mapping from the relative-angle structure of the pair correlation to the final photon momentum dependence through the photon-production kernel. To facilitate future use of these results, the fitting results for the curves in the upper panel of Fig.~\ref{fig:PhotonResult} are given in Appendix~\ref{sec:FitResults}.

\begin{figure}
\centering
\includegraphics[width=1.0\linewidth]{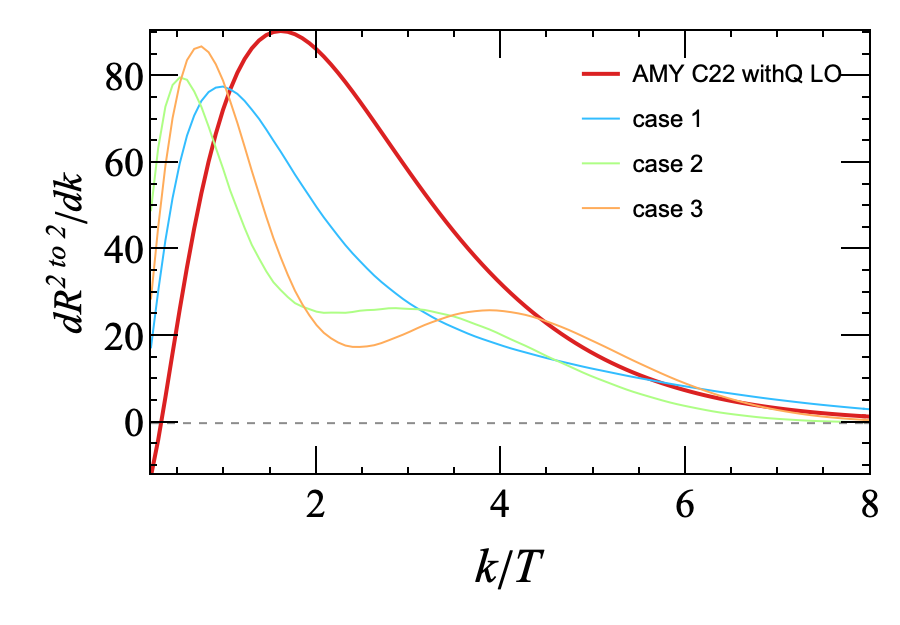}
\caption{Total local photon spectra obtained by adding selected correlation-induced contributions to the AMY $2\to2$ factorized spectrum. The three cases use $(\Psi_{000},\Psi_{001},\Psi_{002},\Psi_{003})=(0,0.5,0,0)$, $(0,0.5,0.2,0)$, and $(0,0.5,0.2,-0.02)$, respectively.}
\label{fig:TotalPhotonResult}
\end{figure}

As a simple illustration of the possible impact on the full local spectrum, we add the correlation-induced contributions to the AMY $2\to2$ factorized result. In Fig.~\ref{fig:TotalPhotonResult}, only the lowest radial sector is used, and the nonzero correlation coefficients are chosen as $(\Psi_{000},\Psi_{001},\Psi_{002},\Psi_{003})=(0,0.5,0,0)$, $(0,0.5,0.2,0)$, and $(0,0.5,0.2,-0.02)$. The resulting spectra show that even a few relative-angle correlation modes can shift the peak position and change the low- and intermediate-momentum shape of the photon spectrum.

\section{Conclusion}\label{sec:Conclusion}

In this work, we have studied how isotropic two-particle correlations in a non-equilibrium QCD medium modify the local in-medium photon-production spectrum. We focused on the $2\to2$ quark--antiquark annihilation and Compton channels, which provide the leading-logarithmic $2\to2$ contribution to photon emission.

To make the calculation compatible with a broad class of two-particle correlation functions, we first decomposed the two-particle distribution into a factorized part and a connected correlation, $\mathcal F_{ab}=f_a f_b+g_{ab}$. We then represented both the single-particle deformation and the pair-correlation profile in a spectral basis. In this representation, the coefficients $\Phi_n$ characterize the single-particle sector, while the coefficients $\Psi_{nm\ell}$ encode the radial and relative-angle structure of the connected two-particle correlation. Different functional forms of $g_{ab}$ can therefore be constructed by choosing different combinations of the basis coefficients $\Psi_{nm\ell}$.

As an illustrative application, we considered a classical-statistics baseline with $\Phi(\xi)=1$, so that changes relative to the factorized contribution arise solely from the connected two-particle sector. 
We found that nonzero relative-angle modes of an isotropic pair correlation generate sign-changing modifications of the photon spectrum, with magnitudes that can be comparable to the factorized contribution. This behavior appears in both the annihilation and Compton channels, while the detailed momentum dependence differs because their scattering kernels probe the correlation structure differently. After summing the annihilation, quark Compton, and antiquark Compton channels, the total $2\to2$ spectrum retains the oscillatory correlation-induced structures. Thus, even though rotational invariance requires the emitted photon distribution to remain isotropic, its energy dependence can still carry information about the relative-angle structure of two-particle correlations in the medium.

The response-function framework developed here provides a basis for several concrete extensions. First, a more quantitative treatment should restore the quantum-statistical factors in the distribution functions and final-state statistical weights. This is necessary because the present numerical study was carried out in the classical-statistics limit, where the low-momentum region is not expected to give a physical photon-production rate. Second, correlation corrections should be incorporated into a complete leading-order photon-production calculation. The present work kept only the leading-logarithmic $2\to2$ contribution, so the nonlogarithmic leading-order terms, together with the collinearly enhanced bremsstrahlung and inelastic pair-annihilation processes required in the full leading-order treatment, remain to be included. Third, we will embed the resulting local rates in realistic space-time evolutions of the collision system. The calculation presented here concerns a local emission rate, while comparison with measured photon spectra requires the integration over the evolving medium. Finally, the same strategy can be applied to other electromagnetic probes, in particular dilepton spectra, where correlations in the medium may provide complementary information about non-equilibrium QCD dynamics.

\section*{Acknowledgments} 

We are grateful to Dr. Pengfei Zhuang for helpful discussions. This work is supported by NSFC by grant No. 12575143 and by Tsinghua University under grant Nos. 04200500123, 531205006, 533305009. We also acknowledge the support by center of high performance computing, Tsinghua University.

\appendix

\section{Response functions and photon-momentum projection}\label{sec:ResponseFunctions}

After the medium distribution functions have been expanded in the Laguerre basis, the remaining photon-momentum dependence can be projected onto the same radial basis. We write the factorized and correlation-induced parts of the momentum-differential photon-production rate as
\begin{align}
\mathcal R_{\chi}^{\mathrm{fac}/\mathrm{corr}}(X;\xi_k)
=
f_\gamma^{\mathrm{ref}}(\xi_k)
\sum_{q=0}^{N_{\gamma}}
L_q^{(2)}(\xi_k)
\mathcal R_{\chi,q}^{\mathrm{fac}/\mathrm{corr}}(X).
\end{align}
Each photon spectral coefficient is obtained from
\begin{align}
\begin{split}
    \mathcal R_{\chi,q}^{\mathrm{fac}/\mathrm{corr}}(X)
=\;&
\frac{q!}{(q+2)!}
\int_0^\infty d\xi_k\,
\xi_k^2 e^{-\xi_k}
\Big[f_\gamma^{\mathrm{ref}}(\xi_k)\Big]^{-1}
\\&\times
L_q^{(2)}(\xi_k)
\mathcal R_{\chi}^{\mathrm{fac}/\mathrm{corr}}(X;\xi_k).
\end{split}
\end{align}
Keeping the terms that appear in the compact rate formula of the main text gives
\begin{align}
\mathcal R_{\chi,q}^{\rm fac}
=
\sum_{n,m}
\Phi_n\Phi_m
\widehat{\mathcal I}^{ab;\chi,0}_{q;nm}
+
s_c
\sum_{n,m,r}
\Phi_n\Phi_m\Phi_r
\widehat{\mathcal I}^{abc;\chi,1}_{q;nmr},
\end{align}
whereas the connected two-particle correlation gives
\begin{align}
\begin{split}
&
\mathcal R_{\chi,q}^{\rm corr}
=\;
\sum_{n,m,\ell}
\Psi_{nm\ell}
\widehat{\mathcal J}^{ab;\chi,0}_{q;nm\ell}
+
s_c
\sum_{n,m,\ell,r}
\Psi_{nm\ell}\Phi_r
\widehat{\mathcal J}^{ab,c;\chi,1}_{q;nm\ell r}
\\&
+
s_c
\sum_{n,m,\ell,r}
\Psi_{nm\ell}\Phi_r
\widehat{\mathcal J}^{ac,b;\chi,1}_{q;nm\ell r}
+
s_c
\sum_{n,m,\ell,r}
\Psi_{nm\ell}\Phi_r
\widehat{\mathcal J}^{bc,a;\chi,1}_{q;nm\ell r}.
\end{split}
\end{align}

The hatted response coefficients are obtained by projecting the photon-momentum differential response functions onto the same Laguerre basis used for the medium. Explicitly,
\begin{align}
\begin{split}
\widehat{\mathcal I}^{ab;\chi,0}_{q;nm}
=\;&
\frac{q!}{(q+2)!}
\int_0^\infty d\xi_k\,
\xi_k^2 e^{-\xi_k}
\Big[f_\gamma^{\mathrm{ref}}(\xi_k)\Big]^{-1}
\\&\times
L_q^{(2)}(\xi_k)
\mathcal I^{ab;\chi,0}_{nm}(\xi_k),
\end{split}
\end{align}
\begin{align}
\begin{split}
\widehat{\mathcal I}^{abc;\chi,1}_{q;nmr}
=\;&
\frac{q!}{(q+2)!}
\int_0^\infty d\xi_k\,
\xi_k^2 e^{-\xi_k}
\Big[f_\gamma^{\mathrm{ref}}(\xi_k)\Big]^{-1}
\\&\times
L_q^{(2)}(\xi_k)
\mathcal I^{abc;\chi,1}_{nmr}(\xi_k),
\end{split}
\end{align}
\begin{align}
\begin{split}
\widehat{\mathcal J}^{ab;\chi,0}_{q;nm\ell}
=\;&
\frac{q!}{(q+2)!}
\int_0^\infty d\xi_k\,
\xi_k^2 e^{-\xi_k}
\Big[f_\gamma^{\mathrm{ref}}(\xi_k)\Big]^{-1}
\\&\times
L_q^{(2)}(\xi_k)
\mathcal J^{ab;\chi,0}_{nm\ell}(\xi_k),
\end{split}
\end{align}
and
\begin{align}
\begin{split}
\widehat{\mathcal J}^{ab,c;\chi,1}_{q;nm\ell r}
=\;&
\frac{q!}{(q+2)!}
\int_0^\infty d\xi_k\,
\xi_k^2 e^{-\xi_k}
\Big[f_\gamma^{\mathrm{ref}}(\xi_k)\Big]^{-1}
\\&\times
L_q^{(2)}(\xi_k)
\mathcal J^{ab,c;\chi,1}_{nm\ell r}(\xi_k).
\end{split}
\end{align}
The projected coefficients $\widehat{\mathcal J}^{ac,b;\chi,1}_{q;nm\ell r}$ and $\widehat{\mathcal J}^{bc,a;\chi,1}_{q;nm\ell r}$ are defined in the same way, with the correlated pair and the spectator leg relabeled accordingly.

The remaining phase-space integrals are evaluated following the method developed in Ref.~\cite{Lu:2025yry}. Since the reduction is purely algebraic and does not introduce additional physical assumptions beyond the response-function definitions given above, we do not reproduce the intermediate steps here. In the numerical implementation, the projected response coefficients
$\widehat{\mathcal I}^{ab;\chi,0}_{q;nm}$,
$\widehat{\mathcal I}^{abc;\chi,1}_{q;nmr}$,
$\widehat{\mathcal J}^{ab;\chi,0}_{q;nm\ell}$, and
$\widehat{\mathcal J}^{ab,c;\chi,1}_{q;nm\ell r}$
are computed separately for each channel $\chi$ and for each set of basis indices. After these coefficients are computed, the full momentum-dependent rate is reconstructed by multiplying them by the corresponding photon basis functions and summing over $q$, with additional sums over the single-particle and two-particle medium coefficients $\Phi_n$ and $\Psi_{nm\ell}$.

\section{Fit Results}\label{sec:FitResults}

This appendix gives analytic fits to the total $2\to2$ correlation-induced photon spectra shown in the upper panel of Fig.~\ref{fig:PhotonResult}. We use the dimensionless photon momentum $\xi_k\equiv k/T$. The fitted curve $\mathcal R^{2\to2,\mathrm{corr}}_{00\ell}(\xi_k)$ denotes the total annihilation plus quark and antiquark Compton contribution generated by the single correlation-basis component $\Psi_{00\ell}$, with all other correlation coefficients set to zero and with $\Phi(\xi)=1$.
\begin{align}
\begin{split}
&
    \mathcal R^{2\to2,\mathrm{corr}}_{000}(\xi_k)
=
    e^{-\xi_k}
    \Big(
    198.913\,\xi_k
    +171.889\,\xi_k\log \xi_k
    \Big),
\end{split}
\end{align}

\begin{align}
\begin{split}
&
    \mathcal R^{2\to2,\mathrm{corr}}_{001}(\xi_k)
=\;
    e^{-\xi_k}
    \Big(
    78.0082\,\xi_k^3
    -525.333\,\xi_k^2
\\&
    +468.756\,\xi_k
    +0.835938\,\xi_k\log \xi_k
    \Big),
\end{split}
\end{align}

\begin{align}
\begin{split}
&
    \mathcal R^{2\to2,\mathrm{corr}}_{002}(\xi_k)
=\;
    e^{-\xi_k}
    \Big(
    12.7198\,\xi_k^5
    -254.233\,\xi_k^4
\\&
    +1490.94\,\xi_k^3
    -3049.35\,\xi_k^2
    +1525.54\,\xi_k
\\&
    -2.09847\,\xi_k\log \xi_k
    \Big),
\end{split}
\end{align}

\begin{align}
\begin{split}
&
    \mathcal R^{2\to2,\mathrm{corr}}_{003}(\xi_k)
=\;
    e^{-\xi_k}
    \Big(
    -4.41437\,\xi_k^6
    +99.7075\,\xi_k^5
\\&
    -779.367\,\xi_k^4
    +2430.89\,\xi_k^3
    -2653.74\,\xi_k^2
    +635.335\,\xi_k
\\&
    -305.591\,\xi_k\log \xi_k
    \Big).
\end{split}
\end{align}

\newpage

\bibliography{biblio.bib}

\end{document}